\documentclass[aps,prd,superscriptaddress,preprint,tightenlines,nofootinbib]{revtex4}

\usepackage{graphicx}
\usepackage{dcolumn}
\usepackage{bm}

\def \b{{\cal B}}
\def \bea{\begin{eqnarray}}
\def \beq{\begin{equation}}

\def \eea{\end{eqnarray}}
\def \eeq{\end{equation}}
\def \jpsi{J/\psi}

\begin{document}

\preprint{CLNS 08/2031}
\preprint{CLEO 08-14}

\title{\boldmath Observation of $\Upsilon(2S) \to \eta \Upsilon(1S)$
and search for related transitions}

\author{Q.~He}
\author{J.~Insler}
\author{H.~Muramatsu}
\author{C.~S.~Park}
\author{E.~H.~Thorndike}
\author{F.~Yang}
\affiliation{University of Rochester, Rochester, New York 14627, USA}
\author{M.~Artuso}
\author{S.~Blusk}
\author{S.~Khalil}
\author{J.~Li}
\author{R.~Mountain}
\author{S.~Nisar}
\author{K.~Randrianarivony}
\author{N.~Sultana}
\author{T.~Skwarnicki}
\author{S.~Stone}
\author{J.~C.~Wang}
\author{L.~M.~Zhang}
\affiliation{Syracuse University, Syracuse, New York 13244, USA}
% ...
\collaboration{CLEO Collaboration}
\noaffiliation

\date{June 17, 2008}

\begin{abstract}
We report the first observation of $\Upsilon(2S) \to \eta
\Upsilon(1S)$, with branching fraction $\b = (2.1^{+0.7}_{-0.6}~({\rm stat.})
\pm 0.3~({\rm syst.})) \times 10^{-4}$ and statistical significance $5.3
\sigma$.  Data were acquired with the CLEO III detector at the CESR $e^+ e^-$
symmetric collider.  This is the first process observed involving
a $b$-quark spin flip.  For related transitions, 90\% confidence
limits in units of $10^{-4}$ are
$\b[\Upsilon(2S) \to \pi^0 \Upsilon(1S)] < 1.8$,
$\b[\Upsilon(3S) \to \eta \Upsilon(1S)] < 1.8$,
$\b[\Upsilon(3S) \to \pi^0 \Upsilon(1S)] < 0.7$,
and $\b[\Upsilon(3S) \to \pi^0 \Upsilon(2S)] < 5.1$.
\end{abstract}

\maketitle

In order to produce a pseudoscalar meson $\eta$ or $\pi^0$ in $\Upsilon(nS) \to
(\eta/\pi^0) \Upsilon(mS)$ transitions, the $b \bar b$ pair must emit either
two M1 (chromomagnetic dipole) gluons or an E1 (chromoelectric dipole) and an
M2 (chromomagnetic quadrupole) gluon \cite{Yan:1980,Kuang:2006me,%
Voloshin:2007dx}, involving the flip of a heavy quark's spin.
In this Letter we present the first observation of $\Upsilon(2S) \to \eta
\Upsilon(1S)$, and a search for similar $\pi^0$ or $\eta$ transitions from the
$\Upsilon(2S)$ and $\Upsilon(3S)$.  A spin-flip of a $b$-quark can shed light
on its chromomagnetic moment, expected to scale as $1/m_b$.  Electromagnetic
transitions involving a $b$-quark spin-flip should also have amplitudes scaling
as $1/m_b$.  They have not previously been observed.

The decay $\psi(2S) \to \eta \jpsi$ was observed in the early days of
charmonium spectroscopy \cite{Tanenbaum:1975gz}.  Its branching fraction is
$\b[\psi(2S) \to \eta \jpsi] = (3.13 \pm 0.08)\%$ \cite{PDG}, while only an
upper limit $\b < 2 \times 10^{-3}$ is known for the corresponding
$\Upsilon(2S) \to \eta \Upsilon(1S)$ process \cite{CUSB2Seta1S}.  The upper
limit for $\Upsilon(3S) \to \eta \Upsilon(1S)$ is $\b < 2.2 \times 10^{-3}$
\cite{Brock91}.  The quark spin-flip involved in $\Upsilon(nS) \to (\eta/\pi^0)
\Upsilon(mS)$ transitions (we consider $3 \ge n > m \ge 1$)
and the $P$-wave nature of the final state imply that
rates should scale from charmonium as $\Gamma \propto (p^*)^3/m_Q^4$
\cite{Yan:1980,Kuang:2006me}, where $p^*$ is the three-momentum of the $\eta$
or $\pi^0$ in the $\Upsilon(nS)$ center-of-mass system and $Q = c,b$ is the
heavy quark.  Hence one expects
\beq \label{eqn:rp}
\frac{\Gamma[\Upsilon(2S,3S) \to
\eta \Upsilon(1S)]}{\Gamma[\psi(2S) \to \eta \jpsi]} = (0.0025,0.0013)~,
\eeq
leading to $\b[\Upsilon(2S,3S) \to \eta \Upsilon(1S)]\simeq (8.0,6.5)\times
10^{-4}$.  Direct calculation in a potential model \cite{Kuang:2006me} yields
$(6.9,5.4) \times 10^{-4}$ for these branching fractions.  All predictions
involve a perturbative calculation of gluon-pair emission followed by a
nonperturbative estimate of materialization of the gluon pair into an $\eta$.
Uncertainties associated with this estimate are difficult to quantify.

Similar predictions can be made for $\pi^0$ transitions under the assumption
that they are due to an isospin-zero admixture in the $\pi^0$.  The
isospin-forbidden decay $\psi(2S) \to \pi^0 \jpsi$ has been seen
\cite{PDG} with a branching fraction of $(1.26 \pm 0.13) \times 10^{-3}$
which is
$(4.03 \pm 0.43)\%$ of that for $\psi(2S) \to \eta \jpsi$.  Using values
of $p^*$ appropriate to each process and assuming the same isospin-zero
admixture in $\pi^0$ governs the transitions $\Upsilon(nS) \to \pi^0
\Upsilon(mS)$, one obtains the scaling predictions
\beq \label{eqn:pzpred}
\frac{\b[\Upsilon(2S,3S) \to \pi^0 \Upsilon(1S)]}
     {\b[\Upsilon(2S,3S) \to \eta  \Upsilon(1S)]}
   = (16 \pm 2,0.42 \pm 0.04)\%~.
\eeq
There is no prediction at present for the kinematically-allowed decay
$\Upsilon(3S) \to \pi^0 \Upsilon(2S)$.

The data in the present analysis were collected in $e^+ e^-$ collisions at the
Cornell Electron Storage Ring (CESR), at center-of-mass energies at and about
30 MeV below the $\Upsilon(2S,3S)$ resonances.  Integrated
luminosities at these resonances were (1.3,1.4) fb$^{-1}$, amounting to
$(9.32 \pm 0.14,5.88 \pm 0.10)$ million decays of
$\Upsilon(2S,3S)$, as in the analysis of Ref.\ \cite{number}.
Events were recorded in the CLEO III detector, equipped with an electromagnetic
calorimeter consisting of 7784 CsI(Tl) crystals and covering 93\% of solid
angle, initially installed in the CLEO II \cite{CLEO2} detector configuration.
The energy resolution of the crystal calorimeter is 5\% (2.2\%) for 0.1 (1) GeV
photons.  The CLEO III tracking system \cite{CLEO3trk} consists of a
silicon strip detector and a large drift chamber, achieving a charged particle
momentum resolution of 0.35\% (1\%) at 1 (5) GeV/$c$ in a 1.5 T axial 
magnetic field.

We look for candidate events of the form $e^+ e^- \to \Upsilon(nS) \to
(\eta/\pi^0)\Upsilon(mS)$ with $\Upsilon(mS) \to \ell^+ \ell^-$, where $\ell =
e,\mu$.  Candidates for $\ell^\pm$ are identified by picking the
two highest-momentum tracks in an event and demanding them to be of opposite
sign. We explore separate $e^+ e^-$ and $\mu^+ \mu^-$ samples in $\Upsilon(mS)$
decays by defining electron candidates to have a high ratio of energy $E$
observed in the calorimeter to momentum $p$ measured in the tracking system,
{\it i.e.}, $E/p > 0.75$, and muon candidates to have $E/p < 0.20$.
We choose lepton candidates from tracks satisfying $|\cos \theta| < 0.83$,
where $\theta$ is the angle with respect to the positron beam direction, to
avoid a region of less uniform acceptance at larger $|\cos \theta|$.  With
these criteria we achieve a very clean separation of electron and muon
candidates.  In order to suppress contributions from Bhabha scattering, we
demand for events with $(\eta,\pi^0) \to \gamma \gamma$ that $e^+$ candidates
satisfy $\cos \theta_{e^+} < 0.5$.  This greatly suppresses Bhabha scattering
background while keeping 93\% of the signal.  Once leptons are identified, the
entire event is kinematically fitted.  We reconstruct the
$\eta$ candidates from their decays to $\gamma \gamma$, $\pi^+ \pi^- \pi^0$,
and $3 \pi^0$.  We did not employ the decay mode $\eta \to \pi^+ \pi^- \gamma$
because of its small branching fraction ($\b = [4.69 \pm 0.10]\%$ \cite{PDG})
and large backgrounds,
primarily from $\Upsilon(nS) \to \pi^+ \pi^- \Upsilon(mS)$.

Photon candidates must be detected in the central region of the
calorimeter ($|\cos \theta| < 0.81$), must not be aligned with the initial
momentum of a track, and should have a lateral shower profile consistent with
that of a photon.  Neutral pion candidates (except in the decay $\eta \to
3 \pi^0$, where we only look for six photon candidates) are reconstructed from
a pair of $\gamma$ candidates
required to have $\gamma \gamma$ invariant mass between 120 and 150 MeV.

Monte Carlo (MC) samples were generated for generic $\Upsilon(2S,3S)$ decays
using the routine \textsc{QQ} \cite{QQ}, and for $\Upsilon(nS) \to (\eta/\pi^0)
\Upsilon(mS)$ and dipion transitions between $\Upsilon$ states using the
package \textsc{EvtGen} \cite{Lange:2001uf}. The final $\Upsilon(mS)$ state was
taken to decay to $e^+ e^-$ or $\mu^+ \mu^-$.  A \textsc{Geant}-based
\cite{GEANT}
detector simulation was used.  These samples, as well as off-resonance
$\Upsilon(2S)$ data, are useful both for validating background suppression
methods and as possible background sources.  In calculating branching fractions
from data, we take $\b[\Upsilon(1S) \to e^+ e^-] = \b[\Upsilon(1S) \to \mu^+
\mu^-] = 0.0248 \pm 0.0005$ \cite{PDG} and $\b[\Upsilon(2S) \to e^+ e^-] =
\b[\Upsilon(2S) \to \mu^+ \mu^-] = 0.0203 \pm 0.0009$ \cite{Adams:2004xa} based
on the
more accurately measured $\mu^+ \mu^-$ branching fractions and assuming lepton
universality.

The $\Upsilon(nS) \to (\eta/\pi^0) \Upsilon(mS)$ MC samples were generated with
$\eta$ and $\pi^0$ decaying through all known decay modes.  These decays
proceed via a $P$-wave, and hence are described by a matrix element
$(${\boldmath$\epsilon$}$_i \times$ {\boldmath$\epsilon$}$^*_f) \cdot 
\mathbf{p}_{(\eta/\pi^0)}$
in the nonrelativistic limit (here $^*$ denotes complex conjugation),
with {\boldmath$\epsilon$}$_{f,i}$ the polarization vectors of the final and
initial $\Upsilon$.  The $\theta$ distribution for the final-state leptons in
$\Upsilon(mS) \to \ell^+ \ell^-$ then is $1 - (1/3) \cos^2 \theta$, and
was used in all signal MC samples for $\Upsilon(nS) \to (\eta/\pi^0)
\Upsilon(mS)$.  For $\Upsilon(nS) \to \pi \pi \Upsilon(mS)$ it was assumed
that the $\Upsilon(mS)$ retains the polarization of the initial $\Upsilon(nS)$,
so the lepton angular distribution for $\Upsilon(mS) \to \ell^+ \ell^-$ is
$1 + \cos^2 \theta$.

As a cross-check, data were analyzed for the known transitions $\Upsilon(nS)\to
\pi \pi \Upsilon(1S)$, and branching fractions were found in sufficiently good
agreement with world averages \cite{PDG}.  We looked for systematic differences
between detection of $\Upsilon(1S)\to e^+ e^-$ and $\Upsilon(1S) \to \mu^+
\mu^-$.  Efficiencies for the two modes can differ as a result of the
requirement on $\cos \theta_{e^+}$ mentioned above.  The branching fractions
calculated from $\Upsilon(1S)\to e^+ e^-$ and $\Upsilon(1S) \to \mu^+ \mu^-$
were found to be equal within statistical uncertainty, and consistent with
those obtained from recoil mass spectra without requiring final leptons.

Kinematic fitting was used to study the decays $\Upsilon(nS) \to (\eta/\pi^0)
\Upsilon(mS)$. The two tracks selected as leptons, including photon
bremsstrahlung candidates within 100 mrad of the initial lepton direction,
were constrained to have the known masses of $\Upsilon(mS)$ with a resultant 
reduced $\chi^2$ ($\chi^2/d.o.f.$), $\chi_R^2  \equiv \chi_{\ell^+ \ell^-,m}^2$
required to be less
than 10.  (For off-resonance data the dilepton masses were reduced by an amount
equal to the initial $M[\Upsilon(nS)]$ minus the off-resonance center-of-mass
energy.)  The sum of the four-momenta of these two fitted tracks, including
photon bremsstrahlung candidates as well as the decay products of the
$\eta/\pi^0$, were further constrained to the initial $\Upsilon(nS)$
four-momentum, with a reduced $\chi_R^2 \equiv \chi_{{\rm EVT},m}^2$
required to be
less than 10, or 3 for $(\eta/\pi^0) \to \gamma \gamma$ to help suppress doubly
radiative Bhabha events.  Some of these Bhabha events can give small
fitted $\chi_{{\rm EVT},m}^2$, but have photon momenta shifted by relatively
large
amounts compared to signal events. To further suppress such events, two-photon
``pull" masses,
defined as (fitted--measured)/$\sigma$, where $\sigma$ is the two-photon mass
resolution, were chosen on the basis of signal MC and off-resonance data
(containing the doubly radiative Bhabha contribution) to lie between --2 and 3.
Over 99\% of the signal MC events for all transitions satisfy this
criterion.  All particles were also required to have common vertices in the
above two constrained fits, with reduced $\chi_{\ell^+ \ell^-,v}^2 < 30$
required for the dilepton vertex, and reduced $\chi_{{\rm EVT},v}^2 < 30$
required for the full event vertex.  

For $\Upsilon(2S) \to \eta \Upsilon(1S)$, the photons from
$\eta \to \gamma \gamma$ have energies $E_\gamma = (281 - 64 \cos
\theta^*)$ MeV, where $\theta^*$ is the angle between the photon in the $\eta$
center-of-mass and the $\eta$ boost, so $217 \le E_\gamma \le 345$ MeV.
Choosing $200 \le E_\gamma \le 360$ MeV then eliminates background from
$\Upsilon(2S) \to \gamma \chi_{bJ} \to \gamma \gamma \Upsilon(1S)$ with little
effect on the $\eta \to \gamma \gamma$ signal.
Using the $\Upsilon(2S) \to \eta \Upsilon(1S)$ MC sample, the $\eta$ candidate
mass distribution was fitted to the sum of a double Gaussian and a linear
background.  Constant background gave a
worse fit because of the
kinematic limit at $M[\Upsilon(2S)] - M[\Upsilon(1S)] = 563$ MeV.  The fitting
range was chosen to be 533 to 563 MeV: roughly symmetric about the $\eta$ peak
($M(\eta) = 547.51 \pm 0.18$ MeV \cite{PDG}) with upper boundary at
$M[\Upsilon(2S)] - M[\Upsilon(1S)]$ above which few events are expected or
observed.  The difference between fits with linear and flat backgrounds was
found to be insignificant compared with the systematic error associated with
fitting range.  The double Gaussian parameters
included a narrow width $\sigma_1 = 0.9$ MeV, a wide width $\sigma_2 = 2.1$
MeV, area of second peak 20\% of total, and mean of second peak 0.14 MeV
below the first.

\begin{figure}
\includegraphics[width=0.48\textwidth]{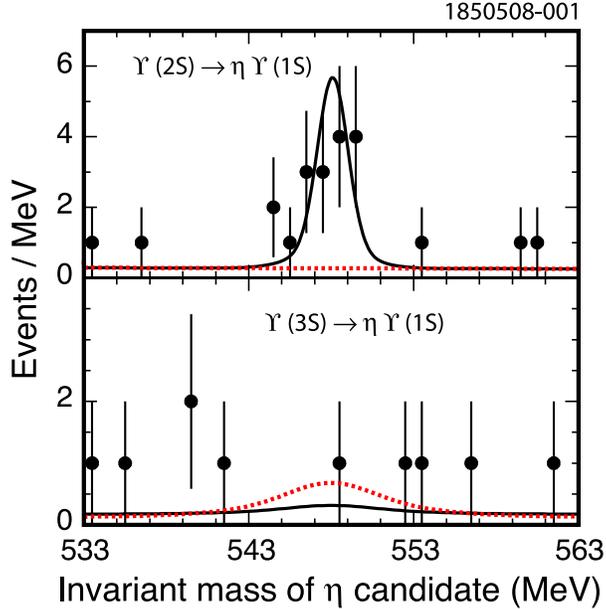}
\caption{Events per MeV vs.\ invariant mass of candidates for $\Upsilon(2S) \to
\eta \Upsilon(1S)$ (top) and $\Upsilon(3S) \to \eta \Upsilon(1S)$ (bottom).
The sum of the modes $\eta \to \gamma \gamma$, $\eta \to \pi^+ \pi^- \pi^0$,
and $\eta \to 3 \pi^0$ is shown.  In the top figure the solid curve corresponds
to the total fit, involving a signal of $13.9^{+4.5}_{-3.8}$ events above
background (dashed line).  In the bottom figure the solid curve corresponds to
a best fit with signal MC shape, while the dotted curve corresponds to a 90\%
confidence level (CL) upper limit.
\label{fig:eta}}
\end{figure}

The mass distribution for the sum of the $\eta$ modes $\gamma\gamma$,
$\pi^+ \pi^- \pi^0$, and $3 \pi^0$ in data (upper plot, Fig.\
\ref{fig:eta}) shows a clear peak near $M(\eta)$.  We fit data
points to the sum of the double Gaussian with floating area but fixed shape
obtained from signal MC and a linear background.  The $\pi^+ \pi^- \pi^0$ and
$3 \pi^0$ decay modes each contribute two events near the peak and none
elsewhere. The combined fitted peak corresponds to a branching fraction
$\b[\Upsilon(2S) \to \eta \Upsilon(1S)] = (2.1^{+0.7}_{-0.6}) \times 10^{-4}$.
Defining the significance $N_\sigma$ as $\sqrt{- 2 \Delta \log{\cal L}}$, where
${\cal L}$ is the likelihood, the difference between fits with and without
signal yields a statistical significance of 5.3 standard deviations.

In searching for $\Upsilon(3S) \to \eta (\to\gamma\gamma) \Upsilon(1S)$
transitions, we suppress backgrounds from cascades involving intermediate
$\chi_b(1P,2P)$ states by requiring one photon to have $500 \le
E_1 \le 725$ MeV and the other to have $140 \le E_2 \le 380$ MeV.
Signal photons satisfy $E_\gamma = (435 - 350 \cos \theta^*)~{\rm MeV}$, so
about 2/3 of them are retained by these choices.  Small differences with
respect to $\Upsilon(2S) \to \eta \Upsilon(1S)$ include (a) an $\eta$ fit range
523--573 MeV
and (b) a flat background, found here to be sufficient to describe
MC and data.  The best fit to signal MC shape and the 90\% confidence level
(CL) upper limit are shown in the lower plot of Fig.\ \ref{fig:eta}.
(No events were observed in the
regions included in the fit but not shown in Fig.\ \ref{fig:eta}.)

\begin{figure}
\includegraphics[width=0.48\textwidth]{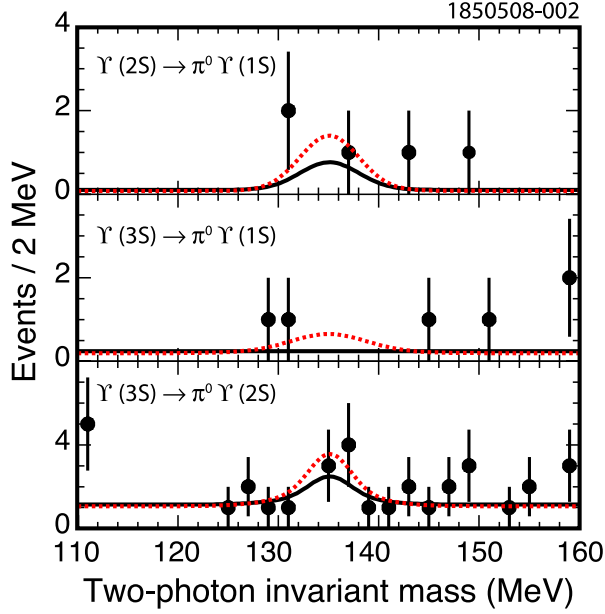}
\caption{Best fits to two-photon invariant mass distributions with signal MC
shapes (solid curves are the results of total fits) and 90\% CL upper limits
(dotted curves) for
$\Upsilon(2S) \to \pi^0 \Upsilon(1S)$ (top), $\Upsilon(3S)\to \pi^0
\Upsilon(1S)$ (middle), and $\Upsilon(3S) \to \pi^0 \Upsilon(2S)$ (bottom).
\label{fig:pi0}}
\end{figure}

For $\Upsilon(2S) \to
\pi^0 \Upsilon(1S)$, the photons from $\pi^0 \to \gamma \gamma$ have energies
$E_\gamma = (274 - 266 \cos \theta^*)$ MeV, so $8 \le E_\gamma \le 540$ MeV.
The choice $200 \le E_\gamma \le 360$ MeV for both photons, made to
eliminate background from $\Upsilon(2S) \to \gamma \chi_{bJ} \to \gamma \gamma
\Upsilon(1S)$, then retains about 30\% of the $\pi^0 \to \gamma \gamma$ signal.
A fit of the $M(\gamma \gamma)$
distribution in the data (using the signal MC
double-Gaussian shape and uniform background) is shown in the top plot of
Fig.\ \ref{fig:pi0}.  Details of this and other limits, as well as of the
$\Upsilon(2S) \to \eta \Upsilon(1S)$ signal, are shown in Table \ref{tab:acc}.
For all $\pi^0$ transitions, MC simulations indicate a constant function is
adequate to describe the background.
Efficiency differences between decay modes are typically due to
details of photon acceptance.

For $\Upsilon(3S) \to \pi^0 \Upsilon(1S)$, where signal photons from $\pi^0
\to \gamma \gamma$ satisfy $E_\gamma = (429 - 385 \cos \theta^*)~{\rm MeV}$,
the same ranges of $(E_1,E_2)$ are chosen as for $\Upsilon(3S) \to \eta
\Upsilon(1S)$.  For $\Upsilon(3S) \to \pi^0 \Upsilon(2S)$, we
suppress backgrounds from cascades involving intermediate $\chi_b(2P)$ states
by {\it excluding} photons with $60 \le E_2 \le 130$ MeV and $190 \le E_1 \le
260$ MeV.  Here, the signal photons satisfy $E_\gamma = (164 - 149 \cos
\theta^*) ~{\rm MeV}$, so about 40\% are retained.  No signal is seen in any of
these $\pi^0$ transitions (Fig.\ \ref{fig:pi0}).

\begin{table}
\caption{Efficiencies, events in data, and product branching fractions $\b
\times \b_\ell$, where $\b \equiv \b[\Upsilon(nS) \to (\eta/ \pi^0)
\Upsilon(mS)]$,
and $\b_\ell\equiv \b[\Upsilon(1S) \to \ell^+ \ell^-] = 4.96\%$ or $\b[\Upsilon
(2S) \to \ell^+ \ell^-]= 4.06\%~(\ell^+ \ell^- \equiv e^+ e^- + \mu^+ \mu^-)$.
Efficiencies are based on MC samples generated with standard $\eta$ and $\pi^0$
branching fractions and with $\b[\Upsilon(mS) \to e^+ e^-]=\b[\Upsilon(mS) \to
\mu^+ \mu^-]=50\%$.  Decays involving $\eta$ are based on combined $\gamma
\gamma$, $\pi^+ \pi^- \pi^0$, and $3\pi^0$ modes.
\label{tab:acc}}
\begin{center}
\begin{tabular}{l c c c c} \hline \hline
\null ~~~~~~ Decay &   MC \%  & Events  & $\b\times\b_\ell$ \\
                   & detected & in data & ($10^{-5}$) \\ \hline
$\Upsilon(2S)$$\to$$\eta \Upsilon(1S)$ & 14.0 &
 $13.9^{+4.5}_{-3.8}$ & $1.06^{+0.35}_{-0.30}$ \\
    $\Upsilon(2S)$$\to$$\pi^0 \Upsilon(1S)$ & 6.8 & $<5.0$ & $<0.79$ \\
    $\Upsilon(3S)$$\to$$\eta\Upsilon(1S)$ & 10.4 & $<4.8$ & $<0.79$ \\
    $\Upsilon(3S)$$\to$$\pi^0 \Upsilon(1S)$ & 13.2 & $<2.3$ & $<0.30$ \\
    $\Upsilon(3S)$$\to$$\pi^0 \Upsilon(2S)$ & 7.8 & $< 8.3$ & $<1.80$ \\
\hline \hline
\end{tabular}
\end{center}
\end{table}

Systematic errors are shown in Table \ref{tab:sys}.  Other contributions
investigated and found to be negligible were (i) cross feeds among $\eta$
modes, (ii) signal shape, (iii) background shape, (iv) triggering details,
and (v) differences in $e/\mu$ reconstruction.
The dominant sources of systematic uncertainties are described below.

{\it 1.  Bhabha event suppression:}  Uncertainties for all processes will arise
from our Bhabha event suppression requirement.  Although it is applied only to
$\gamma \gamma$ modes, it will affect not only $\pi^0$ transitions but also
those with $\eta$, whose $\gamma \gamma$ decays dominate our analyses
statistically.  To probe this uncertainty, 
we consider the separate sample of those events with $\cos \theta_{e^+} \geq
0.5$ which were removed by the Bhabha suppression requirement.  The resultant
$\b[\Upsilon(2S) \to \eta (\to\gamma\gamma) \Upsilon(1S)]$ is consistent
with our nominal result.  Averaging the two gives a deviation of 9\% which we
take as a possible systematic uncertainty due to this requirement.
We then propagate this estimated
uncertainty to the rest of the decay modes with suitable weight for the
fraction of the decay due to $\gamma \gamma$.

{\it 2.  Kinematic fitting:}  To probe any systematic bias
introduced by our kinematic fitting procedure,  we look at events with
very similar topology to our signals:  $\Upsilon(2S)\to\gamma\chi_b(1P_J)$,
$\chi_b(1P_J)\to\gamma\Upsilon(1S)$, $\Upsilon(1S)\to\ell^+ \ell^-$
where $J = 1$ or $2$.  We use the same analysis requirements as for
$\eta\to\gamma\gamma$ but relax the requirements on $E_\gamma$ in
order to accept two-photon cascades through $\chi_b$ states.  Varying
the requirement on $\chi_{{\rm EVT},m}^2$ from \textit{none} to
$\chi_{{\rm EVT},m}^2 < 3$, we observe a maximum deviation
of 7\% in this product of branching fractions which we
assign as a possible source of systematic uncertainty.

\begin{table}
\caption{Systematic errors, in percent, on branching fractions for
$\Upsilon(nS) \to$ (a) $\eta \Upsilon(1S)$; (b) $\pi^0 \Upsilon(1S)$; (c)
$\pi^0 \Upsilon(2S)$.  All errors are assigned symmetrically.  Decays involving
$\eta$ are based on combined $\gamma \gamma$, $\pi^+ \pi^- \pi^0$, and $3\pi^0$
modes. 
The last line (d) includes systematic errors.
\label{tab:sys}}
\begin{center}
{\small
\begin{tabular}{l c c c c c} \hline \hline
~~~~ Decay & \multicolumn{2}{c}{$\Upsilon(2S) \to $} & 
 \multicolumn{3}{c}{$\Upsilon(3S) \to $} \\
Final state & (a) & (b) & (a) & (b) & (c) \\ \hline
Tracks & 2 & 2 & 2 & 2 & 2 \\
Number of $\Upsilon(nS)$ & 1.5 & 1.5 & 1.7 & 1.7 & 1.7 \\
$\eta/\pi^0$ recon.\ & 6 & 5 & 8 & 5 & 5 \\
$\b_{\ell \ell}[\Upsilon(mS)]$ & 2 & 2 & 2 & 2 & 4 \\
$\gamma \gamma$ pull mass & 4 & 0 & 4 & 0 & 0 \\
Bhabha event sup.\ & 7 & 9 & 6 & 9 & 9 \\
Fit range & 1 & 1 & 8 & 6 & 4 \\
$\chi^2$ cuts & 7 & 7 & 7 & 7 & 7 \\
MC stat.\ & 1.1 & 1.6 & 1.3 & 1.1 & 1.5 \\
\hline
Quad.\ sum & 13 & 13 & 16 & 14 & 14 \\ \hline
$\b~(10^{-4})$ (d) & $2.1\!^{+0.7}_{-0.6}$$\pm$0.3 & $<1.8$ & $<1.8$
 & $<0.7$ & $<5.1$ \\
\hline \hline
\end{tabular}
}
\end{center}
\end{table}

{\it 3.  $\eta/\pi^0$ reconstruction:} We assign 5\% per 
$\pi^0$ or $\eta$ decaying into two photons based on 
CLEO studies \cite{pi0syst}.

{\it 4.  Fit ranges:}  Uncertainties due to fit ranges differ
for different final states.  
To estimate them, we prepare many MC samples in which points are randomly
scattered around best-fit values from data (signal plus background),
bin-by-bin according to a Poisson distribution. 
We then fit them with the
fit range boundaries symmetrically changed by $\pm 5$ MeV
for $\Upsilon(2S) \to \eta \Upsilon(1S)$.
In $\Upsilon(3S) \to \eta \Upsilon(1S)$ as well as in
$\Upsilon(nS) \to \pi^0 \Upsilon(mS)$,
where wider kinematic ranges are available,
the fit range boundaries are symmetrically changed by $\pm 10$ MeV.
We assign variations of averages of these fitted yields as
possible systematic shifts.
Combining the effects from the systematic errors linearly with the results
already listed, we find the results shown in the last line of 
Table \ref{tab:sys}.

To summarize, we have observed for the first time a process involving $b$-quark
spin-flip, with 
$\b[\Upsilon(2S) \to \eta \Upsilon(1S)] = (2.1^{+0.7} _{-0.6}
\pm 0.3)\times 10^{-4}$. The statistical significance of the signal is 5.3
$\sigma$.  The result is about 1/4 of the value one would predict on the basis
of Eq.\ (\ref{eqn:rp}), indicating either a shortcoming in the description of
two-gluon hadronization into an $\eta$ or a fundamental suppression of the
chromomagnetic moment of the $b$ quark.  In addition, we have set 90\% CL
upper limits on other pseudoscalar transitions summarized on the bottom line of
Table \ref{tab:sys}.  The limit
on $\b[\Upsilon(3S) \to \eta \Upsilon(1S)]$ is about a factor of two below
that predicted from Eq.\ (\ref{eqn:rp}), while the limits on the transitions
$\Upsilon(2S,3S) \to \pi^0 \Upsilon(1S)$ are consistent with the estimates of
Eq.\ (\ref{eqn:pzpred}).

We gratefully acknowledge the effort of the CESR staff in providing us with
excellent luminosity and running conditions.  This work was supported by the
A. P. Sloan Foundation, the National Science Foundation, the U. S. Department
of Energy, the Natural Sciences and Engineering Research Council of Canada, and
the U. K. Science and Technology Facilities Council.

\end{document}